\documentclass[prd,aps,floats,floatfix,eqsecnum,nofootinbib,showpacs]{revtex4}
\usepackage{verbatim,graphicx,amssymb,amsbsy,bm,amsmath,rotating,epsfig}
\usepackage{psfrag}
\usepackage{hyperref}
\usepackage{fontenc}
\newcommand{\be}{\begin{equation}}
\newcommand{\ee}{\end{equation}}
\newcommand{\bea}{\begin{eqnarray}}
\newcommand{\eea}{\end{eqnarray}}

\newcommand{\bp}{\ensuremath{\mathbf{p}}}

\newcommand{\br}{\ensuremath{\mathbf{r}}}

\begin{document}
\title{Quantum WDM fermions and gravitation determine the observed galaxy structures}
\author{\bf C. Destri $^{(a)}$} \email{Claudio.Destri@mib.infn.it}
\author{\bf H. J. de Vega $^{(b,c)}$}
\email{devega@lpthe.jussieu.fr} 
\author{\bf N. G. Sanchez $^{(c)}$}
\email{Norma.Sanchez@obspm.fr} 
\affiliation{$^{(a)}$ Dipartimento di Fisica G. Occhialini, Universit\`a
Milano-Bicocca and INFN, sezione di Milano-Bicocca, Piazza della Scienza 3,
20126 Milano, Italia. \\
$^{(b)}$ LPTHE, Universit\'e
Pierre et Marie Curie (Paris VI),
Laboratoire Associ\'e au CNRS UMR 7589, Tour 13, 4\`eme. et 5\`eme. \'etages, 
Boite 126, 4, Place Jussieu, 75252 Paris, Cedex 05, France. \\
$^{(c)}$ Observatoire de Paris,
LERMA. Laboratoire Associ\'e au CNRS UMR 8112.
 \\61, Avenue de l'Observatoire, 75014 Paris, France.}
\date{\today}
\begin{abstract}
Quantum mechanics is necessary to compute galaxy structures at kpc scales and below. 
This is so because near the galaxy center, at scales below $ 10 - 100 $ pc,
warm dark matter (WDM) {\bf quantum} effects are important: 
observations show that the interparticle distance is of the order of, or smaller than
the de Broglie wavelength for WDM. This explains why all classical
(non-quantum) WDM $N$-body simulations fail to explain galactic cores and their sizes.
We describe fermionic WDM galaxies in an analytic semiclassical framework based 
on the Thomas-Fermi approach, we resolve it numerically 
and find the main physical galaxy magnitudes: mass, halo radius, phase-space density, 
velocity dispersion, fully consistent with observations, including compact dwarf galaxies.
Namely, fermionic WDM treated quantum mechanically, as it must be, reproduces
the observed galaxy DM cores and their sizes. [In addition, as is known, WDM simulations
produce the right DM structures in agreement with observations for scales $ \gtrsim $ kpc].
We show that compact dwarf galaxies are natural quantum macroscopic objects supported against 
gravity by the fermionic WDM quantum pressure (quantum degenerate fermions) 
with a {\bf minimal} galaxy mass and {\bf minimal} velocity dispersion. 
Interestingly enough, the minimal galaxy mass implies a minimal mass $ m_{min} $ for the WDM particle.
The lightest known dwarf galaxy (Willman I) implies $ m > m_{min} = 1.91 $ keV. 
These results and the observed halo radius and mass of the compact galaxies 
provide further indication that the WDM particle mass $ m $ is approximately around 2 keV.
\end{abstract}
\maketitle
\tableofcontents

\section{Introduction and summary of results}

Dark matter (DM) is the main component of galaxies, especially of dwarf
galaxies which are almost exclusively formed by DM. It thus appears that the study of
galaxy properties is an excellent way to disentangle the nature of DM.

\medskip

Cold DM (CDM) produces an overabundance of substructures below the $ \sim 50 $ kpc till very small scales  
$ \sim 0.005 $ pc which constitutes, as is well known, one of the most serious drawbacks for CDM.
On the contrary, warm DM (WDM),
that is, DM particles with mass in the keV scale, produces DM structures 
in the range of scales $ \lesssim 50 $ kpc in agreement with observations. 
In WDM structure formation, substructures below the free-streaming
scale $ \sim 50 $ kpc are not formed, contrary to the case of CDM.
This conclusion for WDM based on the linear theory is robustely confirmed by
$N$-body simulations by different groups \cite{simuw}. For scales larger than $ ~ 50 $ kpc,
WDM yields the same results than CDM and agrees with all the observations:
small scale as well as large scale structure observations and CMB anisotropy observations.

\medskip

Astronomical observations show that the DM galaxy density profiles are {\bf cored} 
till scales below the kpc \cite{obs,gil,wp}.
On the other hand, $N$-body CDM simulations exhibit cusped density profiles with a typical $ 1/r $ behaviour
near the galaxy center $ r = 0 $. Classical $N$-body WDM simulations exhibit cusps or small
cores smaller than the observed cores \cite{coreswdm,mash}.

\medskip

Numerical calculations based on the spherically symmetric Vlasov--Poisson
     equation from the Larson moment expansion \cite{larson}, as well as on the exact
     dynamics of the associated N-body system, have confirmed these
     findings \cite{nos2}.

\medskip

A direct way to see whether a system of particles has a classical or quantum nature
is to compare the particle de Broglie wavelength $ \lambda_{dB} $ with the inter-particle distance $ d $.
We investigate this issue in sec. \ref{broglie} and express the ratio of the two lengths as 
$$
{\cal R} \equiv \frac{\lambda_{dB}}{d} = \hbar \; \left( \frac{Q_h}{m^4}\right)^{\! \! \frac13} \qquad  ,
\qquad  Q_h \equiv \frac{\rho_h}{\sigma^3} \; .
$$
where $ Q_h $ is the DM phase space density, $ \rho_h $ and $ \sigma $ being the halo density and
the velocity dispersion, respectively. Values of $ \cal R $ much smaller than unity 
correspond to classical physics while
$ {\cal R} \sim 1 $  or larger corresponds to the quantum regime.
The observed  values of $ Q_h $ from Table \ref{pgal} yields $ \cal R $ in the range
\be
2 \times 10^{-3}  < {\cal R} \; \left( \frac{m}{\rm keV}\right)^{\! \!\frac43} < 1.4 \; .
\ee
The larger value of $ \cal R $ is for ultracompact dwarf galaxies and the smaller value of $ \cal R $ 
is for large spirals. The values of $ \cal R $ around unity clearly imply
(and solely from observations) 
that compact dwarf galaxies are natural {\it macroscopic quantum objects} for WDM.

\medskip

WDM fermions always provide a non-zero pressure of quantum nature. By balancing
this quantum pressure with the gravitation pressure, we find theoretical
values for the total mass $ M \sim 10^6 \;  M_\odot$, the radius $ R \sim 30 $ pc and the velocity dispersion
$ \sigma \sim 2 $ km/s which are consistent with the observations of compact dwarf galaxies (see Table \ref{pgal}). 
These results back the idea that dwarf galaxies are supported by the
fermionic {\it WDM quantum pressure}.

\medskip

Classical $N$-body simulations for DM are not valid at scales where the
interparticle distance becomes of the order of, or smaller than, the de Broglie wavelength.
This precisely happens near the galaxy center for WDM particles with mass $ m \sim $ keV.
Therefore, the results of classical (non-quantum) WDM $N$-body simulations at such scales are not valid.
This explains why all classical WDM $N$-body simulations fail to correctly reproduce 
the observed galaxy cores  \cite{mash}. Classical $N$-body simulations
are reliable for CDM because the mass of the CDM particle is large enough 
($ m \sim $ GeV) and then $ {\cal R} \ll 1 $. Such reliable CDM simulations always
produce cusped galaxy profiles in contradiction with the observations.

\medskip

We treat here the self-gravitating fermionic DM in the Thomas-Fermi approximation.
In this approach, the central quantity to derive is the DM chemical potential $ \mu(r) $,
the chemical potential being the free energy per particle \cite{ll}. For self-gravitating
systems the chemical potential is given by $ \mu(r) =  \mu_0 - m \; \phi(r) $,
where $ \mu_0 $ is a constant and $ \phi(r) $ is the gravitational potential. $ \mu(r) $
obeys the self-consistent Poisson equation
\be\label{poisI}
\frac{d^2 \mu}{dr^2} + \frac2{r} \; \frac{d \mu}{dr} = - 
\frac{4 \, \pi \; G \; m^2}{\pi^2 \; \hbar^3} \int_0^{\infty} p^2 \; dp \; f\left(e(p)-\mu(r)\right)
\ee
where $ G $ is Newton's gravitational constant, $ p $ is the DM particle momentum,
$ e(p) = p^2/(2 \, m) $ is the DM particle kinetic energy and $ f(E) $ is the
energy distribution function. This is a semiclassical gravitational approach to determine selfconsistently the
gravitational potential of the fermionic WDM given its distribution function $ f $.

\medskip

The boundary condition $ \mu'(0) = 0 $ at the origin guarantees bounded central DM mass densities.

\medskip

The distribution function $ f(E) $ (or a more general form with a dependence
on other constants of the motion besides $ E $) is determined by the DM evolution since decoupling. 
Such quantum dynamical calculation is beyond the scope of the present paper. 
$ f(E) $ can be modelized for instance by the equilibrium
Fermi-Dirac distributions or by out of equilibrium distributions.
The galaxy magnitudes turn to be rather insensible to whether one chooses
equilibrium or out of equilibrium distributions $ f(E) $ \cite{nos}.

\medskip

We get a one parameter family of solutions of eqs.(\ref{poisI})
parametrized by the value of the chemical potential at the origin  $ \mu(0) $
that can be expressed in terms of the phase-space density at the origin $ Q(0) $. 
Large positive values of $ \mu(0) $ correspond to 
most compact object (fermions in the quantum degenerate limit),
while large negative values of $ \mu(0) $ yield dilute objects (classical limit).

\medskip

We show that the Thomas-Fermi equation implies the local equation of state
$$
P(r) =  \sigma^2(r) \; \rho(r) \quad {\rm and ~ the ~  hydrostatic ~ equilibrium ~  equation}
\quad 
\frac{dP}{dr} + \rho(r) \; \frac{d\phi}{dr} = 0 \; .
$$
This local equation of state generalizes the local perfect fluid equation 
for $r$-dependent velocity $ v(r) $.

\medskip

\begin{figure}[h]
\begin{center}
\includegraphics[width=14.cm]{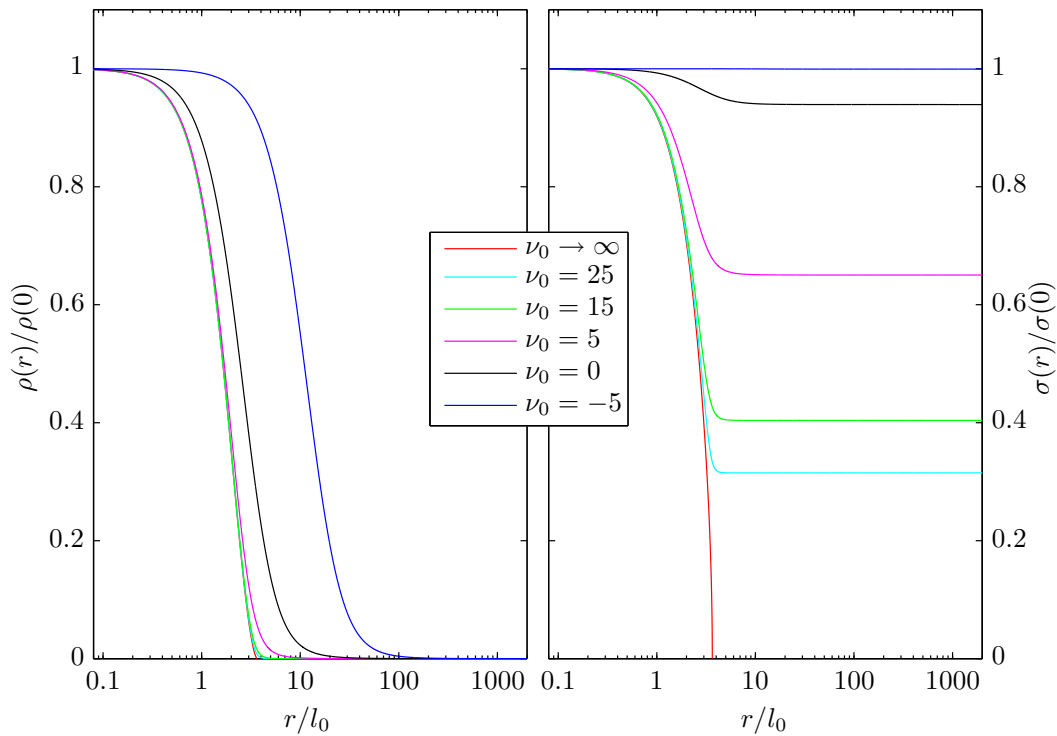}
\caption{Density and velocity profiles, $ \rho(r)/\rho_0 $ and $ \sigma(r)/\sigma(0) $, 
as functions of $ r/l_0 $ for different values of the chemical potential
at the origin $ \nu_0 $. Large positive values of $ \nu_0 $ correspond
to compact galaxies, negative values of $ \nu_0 $ correspond to the classical regime
describing spiral and elliptical galaxies.
All density profiles are cored. The sizes of the cores $ r_h $ defined by eq.(\ref{onequarter}) 
are in agreement with the observations, from the compact galaxies where $ r_h \sim 35 $ pc till
the spiral and elliptical galaxies where $ r_h \sim .2 - 60 $ kpc. The larger and positive is 
$ \nu_0 $, the smaller is the core. The minimal one arises in
the degenerate case  $ \nu_0 \to +\infty $ (compact dwarf galaxies).}
\label{deg}
\end{center}
\end{figure}

The numerical resolution of eqs.(\ref{poisI}) for the whole range of
the chemical potential at the origin  $ \mu(0) $ yields the physical
galaxy magnitudes, such as mass, halo radius, phase-space density and velocity dispersion 
all fully compatible with observations including compact dwarf galaxies as can be
seen from figs. \ref{deg} and \ref{halo} and Table \ref{pgal}.

Approaching the classical diluted limit yields larger and larger halo radii, galaxy masses
and velocity dispersions. Their maximum values are limited by the initial conditions
provided by the primordial power spectrum which determines the sizes and masses of the galaxies formed.
The phase space density decreases from its maximum value for the compact dwarf galaxies 
corresponding to the degenerate fermions limit till its smallest value for large galaxies
(spirals and ellipticals) corresponding to the classical dilute regime.
The theoretical values for the core radius $ r_h $, the core galaxy mass $ M_h $ and 
the velocity dispersion $ \sigma(0) $ obtained in the Thomas-Fermi framework
vary very little with the specific form of the distribution function $ f(E) $.

We display in fig. \ref{deg} the density and velocity profiles, $ \rho(r)/\rho(0) $ and 
$ \sigma(r)/\sigma(0) $ obtained in the Thomas-Fermi approach 
for different values of the chemical potential at the origin $ \mu(0) $. 
Large positive values of $ \mu(0) $ correspond to the compact galaxies, while negative values 
of $ \mu(0) $ correspond to the classical regime describing spiral and elliptical galaxies.
All density profiles are cored. The sizes of the cores $ r_h $
are in agreement with the observations, from the compact galaxies where $ r_h \sim 35 $ pc till
the spiral and elliptical galaxies where $ r_h \sim 0.2 - 60 $ kpc. The larger and positive is 
$ \mu(0) $, the smaller is the core. The minimal one arises in
the degenerate case  $ \mu(0) \to +\infty $ (compact dwarf galaxies).

\medskip

In the left panel of fig. \ref{halo}, we plot the (dimensionless) galaxy phase-space density 
$ \hbar^3 \; Q(0)/({\rm keV})^4 $ obtained from the numerical resolution of the Thomas-Fermi eqs.(\ref{poisI})
for WDM fermions of mass $ m = 1 $ and $ 2 $ keV. The observed values of 
$ \hbar^3 \; Q_h/({\rm keV})^4 $ and $ r_h $ from Table \ref{pgal} are also depicted. 
The theoretical Thomas-Fermi curves in fig. \ref{halo} appear slightly below the observational data
in all the range of galaxies most likely because the observed values $ Q_h $ (derived from the stars'
velocity dispersion) are in fact upper bounds for the DM $ Q_h $. 

In the right panel of fig. \ref{halo}, we plot the galaxy masses 
$ (M_h/M_\odot) \sqrt{M_\odot/{\rm pc}^3 \; \rho_0} $ 
obtained from the numerical resolution of the Thomas-Fermi eqs.(\ref{poisI}),
where $ M_h $ is the mass inside the halo radius and $ \rho_0 $ the observed central mass density.
We observe a good agreement between the Thomas-Fermi results and the observations in {\bf all} the 
range of galaxies for a DM particle mass $ m $ around 2 keV. Notice that the error bars 
of the observational data are not reported here but they are at least about $ 10-20 \%$.

\medskip

For degenerate fermions [$ \mu(0) \to +\infty $] the halo radius, the velocity
dispersion and the galaxy mass take their {\it minimum} values.
These minimum values eqs.(\ref{minimo}) and (\ref{masam}) obtained in the Thomas-Fermi approach
are similar to the estimates provided
in sec. \ref{qup} through a simple balancing argument between the gravitational and quantum pressures. 

The masses of compact dwarf galaxies dominated by DM must be larger than this minimum mass $ M_{h,min} $.
The lightest known  galaxy of this kind is Willman I (see Table \ref{pgal}).
Imposing $ M_{h,min} < M_{Willman ~ I} =  2.9 \; 10^4 \; M_\odot $ 
provides a minimal mass (a lower bound) for the WDM particle:
\be
m > m_{min} = 1.91 \; {\rm keV} \; . 
\ee
This minimal WDM mass is {\it independent} of the WDM particle physics model. $ m_{min} $  is an 
{\bf universal} value irrespective of the shape of the distribution function $ f(E) $ in the non-degenerate regime.

\medskip

Interestingly enough, comparison of the theoretically derived galaxy masses with the galaxy 
data plotted in fig. \ref{halo} indicates a WDM particle mass $ m $ approximately around  2 keV
in agreement with earlier estimations \cite{dvs,dvss}. 
If the WDM particle mass would be $ m \gg 1 $ keV, an overabundance of small galaxies 
(small scale structures) without observable counterpart would appear.

\medskip

In summary, the theoretical Thomas-Fermi results are fully consistent with all the observations including
 dwarf compact galaxies as can be seen from figs. \ref{deg} and \ref{halo}.
It is highly remarkable that in the context of fermionic WDM the simple stationary
quantum description provided by this semiclassical framework
is able to reproduce such broad variety of galaxies.

\medskip

These results indicate that fermionic WDM treated quantum mechanically (even approximately)
is fully consistent with the observed galaxy properties including the DM core sizes.
That is, quantum physics and the associated quantum pressure, 
{\bf rule out} galaxy cusps for fermionic WDM and provide the right sized observed cores.

\medskip

\begin{table}
\begin{tabular}{|c|c|c|c|c|c|} \hline  
 & & & & & \\
 Galaxy  & $ \displaystyle \frac{r_h}{\rm pc} $ & $  \displaystyle \frac{\sigma}{\frac{\rm km}{\rm s}} $
& $ \displaystyle  \frac{\hbar^{\frac32} \;\sqrt{Q_h}}{({\rm keV})^2} $ & 
$ \rho(0)/\displaystyle \frac{M_\odot}{({\rm pc})^3} $ & $ \displaystyle \frac{M_h}{10^6 \; M_\odot} $
\\ & & & & & \\ \hline 
Willman 1 & 19 & $ 4 $ & $ 0.85 $ & $ 6.3 $ & $ 0.029 $
\\ \hline  
 Segue 1 & 48 & $ 4 $ & $ 1.3 $ & $ 2.5 $ & $ 1.93 $ \\ \hline  
  Leo IV & 400 & $ 3.3 $ & $ 0.2 $ & $ .19 $ & $ 200 $ \\ \hline  
Canis Venatici II & 245 & $ 4.6 $ & $ 0.2 $   & $ 0.49 $ & $ 4.8 $
\\ \hline  
Coma-Berenices & 123 & $ 4.6 $  & $ 0.42 $   & $ 2.09 $  & $ 0.14 $
\\ \hline  
 Leo II & 320 & $ 6.6 $ & $ 0.093 $  & $ 0.34 $ & $ 36.6 $
\\  \hline  
 Leo T & 170 & $ 7.8 $ &  $ 0.12 $  & $ 0.79 $ & $ 12.9 $
\\ \hline  
 Hercules & 387 & $ 5.1 $ &  $ 0.078 $  & $ 0.1 $ & $ 25.1 $
\\ \hline  
 Carina & 424 & $ 6.4 $ & $ 0.075 $  & $ 0.15 $ & $ 32.2 $
\\ \hline 
 Ursa Major I & 504 & 7.6  &  $ 0.066 $  & $ 0.25 $ & $ 33.2 $
\\ \hline  
 Draco & 305 & $ 10.1 $ &  $ 0.06 $  & $ 0.5 $ & $ 26.5 $
\\ \hline  
 Leo I & 518  & $ 9 $ &  $ 0.048 $  & $ 0.22 $ & $ 96 $
\\ \hline  
 Sculptor & 480  & $ 9 $ & $ 0.05 $  & $ 0.25  $ & $ 78.8 $
\\ \hline 
 Bo\"otes I & 362 & $ 9 $ & $ 0.058 $  & $ 0.38 $ & $ 43.2 $
\\ \hline  
 Canis Venatici I & 1220  & $ 7.6 $ & $ 0.037 $ & $ 0.08 $ & $ 344 $
\\ \hline  
Sextans & 1290 & $ 7.1 $ & $ 0.021 $ & $ 0.02 $ & $ 116 $
\\ \hline 
 Ursa Minor & 750 & $ 11.5 $ & $ 0.028 $  & $ 0.16 $ & $ 193 $
\\ \hline  
 Fornax  & 1730 & $ 10.7 $ & $ 0.016 $  & $ 0.053  $ & $ 1750 $
\\  \hline  
 NGC 185  & 450 & $ 31 $ & $ 0.033 $ & $ 4.09 $ & $ 975 $
\\ \hline  
 NGC 855  & 1063 & $ 58 $ & $ 0.01 $ & $ 2.64 $ & $ 8340 $
\\ \hline  
  Small Spiral  & 5100  & $ 40.7 $ & $ 0.0018 $ & $ 0.029 $ & $ 6900 $
\\ \hline  
NGC 4478 & 1890 & $ 147 $ & $ 0.003 $ & $ 3.7 $ & $ 6.55 \times 10^4 $
\\ \hline  
 Medium Spiral & $ 1.9 \times 10^4 $ & $ 76.2 $ & $ 3.7 \times 10^{-4} $ & $ 0.0076 $ & $ 1.01 \times 10^5 $
\\ \hline  
 NGC 731 & 6160 & $ 163 $ & $ 9.27 \times 10^{-4} $ & $ 0.47 $ & $ 2.87 \times 10^5 $
\\ \hline 
 NGC 3853   & 5220 & $ 198 $ & $ 8.8 \times 10^{-4} $ & $ 0.77 $  
& $ 2.87 \times 10^5 $ \\ \hline 
NGC 499  & 7700 &  $ 274 $ & $ 5.9 \times 10^{-4} $ & 
$ 0.91 $ & $ 1.09 \times 10^6 $ \\   \hline 
Large Spiral & $ 5.9 \times 10^4 $ & $ 125 $ & $ 0.96 \times 10^{-4} $ & $ 2.3 \times 10^{-3} $ & 
$ 1. \times 10^6 $ \\ \hline  
\end{tabular}
\caption{Observed values $ r_h, \; \sigma, \; \sqrt{Q_h}, \; \rho(0)$ 
and $ M_h $ covering from ultracompact galaxies to large spiral galaxies
from refs.\cite{wp,sltg,dvss,gil,jdsmg,simon11,wolf10,datos,martinez}. The phase space density is larger
for smaller galaxies, both in mass and size.
Notice that the phase space density is obtained
from the stars velocity dispersion which is expected to be smaller than the DM  velocity dispersion.
Therefore, the reported $ Q_h $ are in fact upper bounds to the true values \cite{jdsmg}.}
\label{pgal}
\end{table}

\medskip

We have not considered baryons in the present analysis of the galaxies.
This is fully justified for dwarf compact galaxies which are composed today
99.99\% of dark matter \cite{barena,datos,mwal}. In large galaxies the baryon fraction can
reach values up to  1 - 3 \% \cite{bariones}.
We have also ignored supermassive central black holes which
appear in large spiral galaxies.
In any case, it must be noticed that the central black hole mass is at most 
$ \sim 10^{-3} $ of the mass of the bulge.

Fermionic WDM by itself produce galaxies and structures in agreement with observations.
Therefore, the effect of including baryons is expected to be a correction to the pure WDM results,
consistent with the fact that dark matter is in average six times more abundant than baryons.

We use units such that the speed of light is $ c = 1 $ throughout this paper.

\subsection{Dwarf galaxies as WDM quantum macroscopic objects}\label{broglie}

In order to determine whether a system of particles has a classical or quantum nature
we should compare the particle de Broglie wavelength with the interparticle distance.

\medskip

The de Broglie wavelength of DM particles in a galaxy can be expressed as
\be\label{LdB}
\lambda_{dB}  = \frac{\hbar}{m \; \sigma} \; ,
\ee

where $ \sigma $ is the velocity dispersion, while the average interparticle distance $ d $ can be estimated as
\be\label{dis}
d = \left( \frac{m}{\rho_h} \right)^{\! \! \frac13} \; ,
\ee
where $ \rho_h $ is the average density in the  galaxy core.
We can measure the classical or quantum character of the system by considering the ratio
$$ 
{\cal R} \equiv \frac{\lambda_{dB}}{d}
$$
By using the phase-space density
$$
 Q_h \equiv \frac{\rho_h}{\sigma^3}
$$
and eqs.(\ref{LdB})-(\ref{dis}), $ \cal R $ can be expressed as
\be
{\cal R} = \hbar \; \left( \frac{Q_h}{m^4}\right)^{\! \! \frac13} \; .
\ee
Notice that $ \cal R $ as well as  $ Q_h $ are invariant under the expansion of the universe
because both $ \lambda_{dB} $ and $ d $ scale with the expansion scale factor.  $ \cal R $ and $ Q_h $
evolve by nonlinear gravitational relaxation.

Using now the observed  values of $ Q_h $ from Table \ref{pgal} yields $ \cal R $ in the range
\be
2 \times 10^{-3}  < {\cal R} \; \left( \frac{m}{\rm keV}\right)^{\! \! \frac43} < 1.4
\ee
The larger value of $ \cal R $ is for ultracompact dwarfs while the smaller value of $ \cal R $ 
is for big spirals.

\medskip

The ratio $ \cal R $ around unity clearly implies a macroscopic quantum object.
Notice that $ \cal R $ expresses solely in terms of $ Q $ and hence 
$ (\hbar^3 \; Q/m^4) $ measures how quantum or classical is the system, here, the galaxy. 
Therefore, we conclude {\bf solely from observations} 
that compact dwarf galaxies are natural macroscopic quantum objects for WDM.

\subsection{Dwarf Galaxies supported by WDM fermionic quantum pressure}\label{qup}

For an order--of--magnitude estimate, let us consider a halo of mass $ M $ and radius 
$ R $ of fermionic matter. Each fermion can be considered inside
a  cell of size $ \Delta x \sim 1 / n^{\frac13} $ and therefore has a momentum
$$
p \sim \frac{\hbar}{\Delta x} \sim \hbar \; n^{\frac13} \; .
$$
The associated quantum pressure $ P_q $ (flux of the momentum) has the value 
\be\label{presq}
P_q = n \;\sigma\; p \sim \hbar \;\sigma\; n^{\frac43} = \frac{\hbar^2}{m} \; n^{\frac53} \; .
\ee
where $ \sigma $ is the dispersion velocity estimated as
$$
\sigma = \frac{p}{m} = \frac{\hbar}{m} \; n^{\frac13} \; .
$$ 
We estimate the number density as
$$
n = \frac{M}{\frac43 \; \pi \; R^3 \; m} \; ,
$$
where $ R $ is an estimation of the halo radius
and we use that $ p = m \;\sigma$ to obtain from eq.(\ref{presq})
\be\label{pqu}
 P_q = \frac{\hbar^2}{m \; R^5} \; \left(\frac{3 \; M}{4 \; \pi \; m}\right)^{\! \! \frac53} \; .
\ee
On the other hand, as is well known, galaxy formation as all structure formation in the Universe
is driven by gravitational physics.
The system will be in dynamical equilibrium if this quantum pressure is balanced by
the gravitational pressure
\be\label{pgr}
P_G = {\rm gravitational~ force}/{\rm area} = \frac{G \; M^2}{R^2} \times \frac1{4 \, \pi \; R^2}
\ee
Equating $  P_q = P_G $ from eqs.(\ref{pgr})-(\ref{pqu})
yields the following expressions for the size $ R $ and the velocity $\sigma$ 
in terms of the mass $ M $ of the system and the mass $ m $ of the particles,:
\bea\label{estM}
&& R = \frac{3^\frac53}{(4 \; \pi)^\frac23} \;
\frac{\hbar^2}{G \; m^\frac83 \; M^\frac13} = 10.61 \; 
\left(\frac{10^6 \;  M_\odot}{ M}\right)^{\! \! \frac13} \; 
\left(\frac{\rm keV}{m}\right)^{\! \! \frac83} \; {\rm pc} \; , \\ \cr
&&\sigma= \left(\frac{4 \, \pi}{81}\right)^{\! \! \frac13} \; \frac{G}{\hbar} \;  m^\frac43 \; M^\frac23=
11.62 \; \left(\frac{m}{\rm keV}\right)^{\! \! \frac43} \; 
\left(\frac{M}{10^6 \;  M_\odot}\right)^{\! \! \frac23} \; \frac{\rm km}{\rm s} \; .
\eea
Notice that the values of $ M , \;  R $ and $ \sigma $ are consistent with the observed values of
dwarf galaxies. Namely, for $ M $ of the order $ 10^6 \;  M_\odot $
(which is a typical mass value for dwarf galaxies), $ R $ and  $\sigma$ 
give the correct order of magnitude for the size and velocity dispersion of dwarf galaxies 
(see Table \ref{pgal}) for a WDM particles mass in the keV scale.

These results back the idea that dwarf galaxies are supported by the
fermionic {\it WDM quantum pressure} eq.(\ref{pqu}).

It is useful to express the above quantities in terms of the density $ \rho $, as
follows
\bea\label{semic}
&& M = \frac{9 \; \hbar^3}{2 \; m^4} \; \sqrt{\frac{\rho}{\pi \; G^3}} =  0.7075 \; 10^5 \; M_\odot \;
\sqrt{\rho \; \frac{{\rm pc}^3}{ M_\odot}} \; \left(\frac{\rm keV}{m}\right)^{\! \! 4}  \; M_\odot \; \, ,  \cr \cr
&& R = \frac{3 \; \hbar}{2 \; \sqrt{\pi \; G}} \; \frac1{m^\frac43 \; \rho^\frac16} = 
25.66 \; \left(\frac{M_\odot}{\rho \; {\rm pc}^3 }\right)^{\! \! \frac16} 
\, \left(\frac{\rm keV}{m}\right)^{\! \! \frac43} \; {\rm pc} \; , \cr \cr
&&\sigma= \hbar \; \left( \frac{\rho}{m^4}  \right)^{\! \! \frac13} = 1.988 
\; \left( \rho \; \displaystyle \frac{{\rm pc}^3 }{M_\odot}\right)^{\! \! \frac13} \;
\left(\frac{\rm keV}{m}\right)^{\! \! \frac43} \; \frac{{\rm km}}{{\rm s}}\, ,  
 \cr \cr
&& P_q = \hbar^2 \; \frac{\rho^\frac53}{m^\frac83}  = 0.04399
\; \left( \rho \; \displaystyle \frac{{\rm pc}^3 }{M_\odot}\right)^{\! \! \frac53} \;
 \left(\frac{\rm keV}{m}\right)^{\! \! \frac83}  \;  \frac{M_\odot}{{\rm kpc}^3 } \label{vsemic} \, .
\eea
$ R $ and $ M $ are typical {\it semiclassical gravitational} quantities involving both $ G $ and
$ \hbar $. The particle velocity $ \sigma $ and the pressure $ P_q $ are of purely quantum 
mechanical origin.

\section{General Galaxy properties from quantum  fermionic WDM in the Thomas-Fermi approach}\label{tf}

We consider a single DM halo in the late stages of structure formation when DM
particles composing it are non--relativistic and their phase--space distribution
function $ f(t, \br,\bp) $ is relaxing to a time--independent form, at least for
$\br$ not too far from the halo center. In the Thomas--Fermi approach such a
time--independent form is taken to be a energy distribution function $f(E)$ of
the conserved single--particle energy $E = p^2/(2m) - \mu $, where $m$ is the
mass of the DM particle and $\mu$ is the chemical potential
\be \label{potq}
  \mu(\br) =  \mu_0 - m \, \phi(\br) 
\ee
with $\phi(\br)$ the gravitational potential and  $ \mu_0 $ some constant.

We consider the spherical symmetric case where
the Poisson equation for $\phi(r)$ takes the form
\be \label{pois}
  \frac{d^2 \mu}{dr^2} + \frac2{r} \; \frac{d \mu}{dr} = - 4\pi \, G \, m \, \rho(r)\; , 
\ee
where $ G $ is Newton's constant and $ \rho(r) $ is the DM mass density. In
turn, $ \rho(r) $ is expressed here as a function of $\mu(r)$ through the
standard integral of the DM phase--space distribution function over the momentum
\be \label{den}
  \rho(r) = \frac{g \, m}{2 \, \pi^2 \, \hbar^3} \int_0^{\infty} dp\;p^2 
  \; f\left(\displaystyle \frac{p^2}{2m}-\mu(r)\right)\; , 
\ee
where $ g $ is the number of internal degrees of freedom of the DM particle,
with $ g = 1 $ for Majorana fermions and $ g = 2 $ for Dirac fermions. For
definiteness, we will take $g=2$ in the sequel.

\medskip

Another standard integral of the DM phase--space distribution function is the pressure
\be \label{P}
  P(r) = \frac{1}{3 \, \pi^2 \,m\,\hbar^3} \int_0^{\infty} dp\;p^4 
  \,f\left(\displaystyle \frac{p^2}{2m}-\mu(r)\right)
\ee
and from $ \rho(r) $ and $ P(r) $ other quantities of interest, such as the velocity dispersion
$ \sigma(r) $ and the phase--space density $ Q(r) $ can be determined as
\be \label{sigQ}
  \sigma^2(r) = \frac{P(r)}{\rho(r)} \quad,\qquad Q(r) = \frac{\rho(r)}{\sigma^3(r)}  \; .
\ee
We see that $\mu(r)$ fully characterizes the fermionic DM halo in this
Thomas--Fermi framework. The chemical potential is monotonically decreasing in $ r $ 
since eq.~(\ref{pois}) implies
\be\label{dmu}
\frac{d\mu}{dr} = -\frac{G\,m\,M(r)}{r^2} \quad,\qquad  
  M(r) = 4\pi \int_0^r dr'\, r'^2 \, \rho(r') \; .
\ee
Moreover, the fermionic DM mass density $ \rho$ is bounded at the origin due to the Pauli principle \cite{nos},
and therefore the proper boundary condition at the origin is
\be
  \frac{d \mu}{dr}(0) = 0 \; .
\ee
Also notice from eq.(\ref{potq}) that, by assuming $ \phi(0)=0 $, then $ \mu_0 \equiv \mu(0) $,
which plays the role of a Lagrange multiplier to be eventually eliminated in favor of
$ \rho_0 \equiv \rho(0) $, the mass density at the origin. 

\medskip

Eqs.(\ref{pois}) and (\ref{den}) provide an ordinary nonlinear
differential equation that determines selfconsistently the chemical potential $ \mu(r) $ and
constitutes the Thomas--Fermi semi-classical approach. Fermionic DM in
this approach has been previously considered in ref. \cite{peter}.

In this semi-classical framework the stationary energy distribution function $
f(E) $ must be assigned beforehand. In a full--fledged treatment one would
rather solve the cosmological DM evolution since decoupling till today, including the quantum
dynamical effects which become important in the non-linear stage and close
enough to the origin. Such a quantum dynamical calculation is beyond the scope
of the present paper.  

\medskip

From eq.(\ref{P}) and (\ref{sigQ}) we derive the local equation of state:
\be\label{eces}
P(r) =  \sigma^2(r) \; \rho(r) \;  \; ,
\ee
and the hydrostatic equilibrium equation
\be\label{ehidr}
\frac{dP}{dr} + \rho(r) \; \frac{d\phi}{dr} = 0 \; .
\ee
The local equation of state eq.(\ref{eces}) generalizes the local perfect fluid equation of state 
for $r$-dependent velocity $ v(r) $.
As we see below, the perfect fluid equation of state
is recovered both in the classical dilute limit and in the quantum degenerate limit.

Eliminating $ P(r) $ between eqs.(\ref{eces}) and (\ref{ehidr}) and integrating on $ r $ gives
$$
\frac{\rho(r)}{\rho(0)} = \frac{\sigma^2(0)}{\sigma^2(r)} \; 
e^{ \displaystyle -\int_0^r\frac{dr'}{\sigma^2(r')} \; \frac{d\phi}{dr'}} \; .
$$
For constant $ v(r) $ this relation reduces to the baryotropic equation.
Inserting this expression for  $ \rho(r) $ in the Poisson's equation yields
$$
\frac{d^2 \phi}{dr^2} + \frac2{r} \; \frac{d \phi}{dr} =  4 \, \pi \; G \; m \; \rho_0 \;
\frac{\sigma^2(0)}{\sigma^2(r)} \; e^{- \displaystyle \int_0^r\frac{dr'}{\sigma^2(r')} \; 
\frac{d\phi}{dr'}} \; .
$$
This nonlinear equation generalizes the corresponding equation in the self-gravitating Boltzmann gas
when $ \sigma^2(r) $ is constant.

\medskip

We integrate the Thomas-Fermi nonlinear differential
equations (\ref{pois})-(\ref{den}) from $ r = 0 $ till the 
boundary $ r = R = R _{200} \sim R_{vir} $ defined as the radius where the 
mass density equals $ 200 $ times the mean DM density.
It is useful to define dimensionless variables $ \xi , \; \xi_R , \; \nu(\xi) $ 
and the dimensionless  one--parameter distribution function $ \Psi $ as
\be\label{varsd}
 r = l_0 \; \xi \qquad , \qquad R = l_0 \; \xi_R \qquad , \qquad 
\mu(r) =  E_0 \;  \nu(\xi) \qquad , \qquad f(E) = \Psi \! \left(\frac{E}{ E_0}\right)
  \; ,
\ee
where $ E_0 $ is the characteristic one--particle energy of the DM halo and 
$ l_0 $ is the characteristic length that emerges from the dynamical equations 
(\ref{pois})-(\ref{den}):
\be\label{varsd2}
l_0 \equiv  \frac{\hbar}{\sqrt{8\,G}} \left(\frac{9\pi}{m^8\,\rho_0}\right)^{\! \! \frac16} 
  = R_0 \; \left(\frac{{\rm keV}}{m}\right)^{\! \! \frac43}  \; 
  \left(\rho_0 \; \frac{{\rm pc}^3}{M_\odot}\right)^{\! \! -\frac16} 
  \;,\qquad R_0 = 18.71 \; \rm pc  \; ,
\ee
The self-consistent Thomas-Fermi equation (\ref{pois})-(\ref{den}) for $ \nu(\xi) $ takes the form
\be\label{nu}
\frac{d^2 \nu}{d\xi^2} + \frac2{\xi} \; \frac{d \nu}{d\xi} = - \frac{I_2(\nu)}{[I_2(\nu_0)]^{1/3}}
\quad ,  \quad \rho(\xi_R) = 200 \; {\bar \rho}_{DM} \quad ,  \quad
\nu'(0) = 0 \quad ,  \quad \nu_0 \equiv \nu(0) \quad  , 
\ee
where
\be\label{dfI}
I_n(\nu) \equiv (n+1) \; \int_0^{\infty} y^n \; dy \; \Psi(y^2 -\nu) \; , \; n = 1, 2 , \ldots\; ,
\ee
and we use the integration variable $ y \equiv p / \sqrt{2 \, m \;  E_0} $.

\medskip

We find the main physical galaxy magnitudes, such as the
mass density $ \rho(r) $, the average velocity of the particles $ v(r) $
and the pressure $ P(r) $ (which are all $r$-dependent) as: 
\bea\label{gorda}
&& \rho(r) = \rho_0 \; \frac{I_2(\nu(\xi))}{I_2(\nu_0)} \quad , \quad
\sigma^2(r) =\sigma^2(0) \;  \frac{I_2(\nu_0)}{I_4(\nu_0)} \frac{I_4(\nu(\xi))}{I_2(\nu(\xi))}
\quad , \quad  P(r) =  P(0)\; \frac{I_4(\nu(\xi))}{I_4(\nu_0)}   \;, \\ \cr
&& \rho_0 = \frac{m^4}{3 \; \pi^2\;\hbar^3}\left(\frac{2\; E_0}{m}\right)^{3/2}
  I_2(\nu_0) \;,\quad 
  P(0) = \frac{2 \; E_0}{5\,m} \; \rho_0 \; \frac{I_4(\nu_0)}{I_2(\nu_0)}
  =  \frac{\hbar^2}{5} \; \left(\frac{3\pi^2}{m^4}\right)^{\! \! \frac23}
  \left[\frac{\rho_0}{I_2(\nu_0)}\right]^{5/3} I_4(\nu_0)\; , \cr \cr
&& \sigma(0) =  V_0 \; \frac{[I_4(\nu_0)]^{1/2}}{[I_2(\nu_0)]^{5/6}} \;
  \left(\frac{{\rm keV}}{m}\right)^{4/3}
\left(\rho_0 \; \frac{{\rm pc}^3}{M_\odot}\right)^{1/3}  
\quad  , \quad  V_0 = 2.751 \; \frac{\rm km}{\rm s} \; .
\eea
As a consequence, the total mass $ M_R $ enclosed in the sphere of radius $ R $ and
the phase space density $ Q(r) $ turn to be
\bea\label{cero} 
&& M_R = 4 \, \pi \int_0^R r^2 \; dr \; \rho(r) =  
4 \, \pi \; \frac{\rho_0\; l_0^3}{I_2(\nu_0)}\,\int_0^{\xi_R}
    dx\, x^2 \,I_2(\nu(x)) = 4 \, \pi \; \frac{\rho_0 \; l_0^3}{[I_2(\nu_0)]^{2/3}} \;
    \xi_R^2 \; |\nu'(\xi_R)| = \cr \cr\cr
&&=  M_0 \; \frac{\xi_R^2 \; |\nu'(\xi_R)|}{[I_2(\nu_0)]^{2/3}}
    \; \left(\frac{{\rm keV}}{m}\right)^{\! \! 4} \; \sqrt{\rho_0 \; \frac{{\rm pc}^3}{M_\odot}} \; , 
\quad  M_0 = 4 \; \pi \; M_\odot \; \left(\frac{R_0}{\rm pc}\right)^{\! \! 3} 
    = 0.8230 \; 10^5 \; M_\odot \; ,\\ \cr 
&& Q(r) = Q(0) \; \frac{I_2^{\frac{5}{2}}(\nu(\xi))}{I_4^{\frac32}(\nu(\xi))}
\; \frac{I_4^{\frac32}(\nu_0)}{I^{\frac52}_2(\nu_0)}
\quad , \quad Q(0) = \frac{\sqrt{125}}{3 \; \pi^2 \; \hbar^3} \; \; m^4 \; \; 
\frac{I^{\frac52}_2(\nu_0)}{I_4^{\frac32}(\nu_0)} \quad  , \quad 
\nu_0 \equiv \nu(0) \quad  , \quad \rho_0 = \rho(0) \; .
\eea
We have systematically eliminated the energy scale $ E_0 $ 
in terms of the central density $ \rho_0 $. [We may also choose the density at another point $ r \neq 0 $].
Notice that $ Q(r) $ turns to be independent of $ E_0 $ and therefore from $ \rho_0 $.

\medskip

Besides the virial galaxy radius $ R = l_0 \; \xi_R $, one can 
define the core size $ r_h $ of the halo by analogy with the Burkert density profile as
\be\label{onequarter}
  \frac{\rho(r_h)}{\rho_0} = \frac14 \quad , \quad  r_h = l_0 \; \xi_h \; .
\ee
To explicitly solve eq. (\ref{nu}) we need to specify the distribution function
$ \Psi(E/E_0) $. But many important properties of the Thomas--Fermi semi-classical
approximation do not depend on the detailed form of the distribution function
$ \Psi(E/E_0) $. Indeed, a generic
feature of a physically sensible one--parameter form $ \Psi(E/E_0) $ is that it should
describe degenerate fermions for $ E_0 \to 0 $. That is, $ \Psi(E/E_0) $ should behave as
the step function $ \theta(-E) $ in such limit. In the opposite limit, $ \Psi(E/E_0) $ 
describes classical particles for $ \mu/E_0 \to -\infty $.
As an example of distribution function, we consider the Fermi--Dirac distribution 
\be\label{FD}
  \Psi_{\rm FD}(E/E_0) = \frac1{e^{E/E_0} + 1} \; .
\ee
The choice of $ \Psi_{\rm FD} $ may be justified near the
origin, where relaxation to thermal equilibrium is conceivable. Far from the  origin
however, the Fermi--Dirac distribution as its classical counterpart, the isothermal sphere,
produces a mass density tail $ 1/r^2 $ that overestimates the observed tails of the 
galaxy mass densities. Notice indeed that the
classical regime $ \mu/E_0 \to -\infty $ may be attained for large distances $ r $
since eq.(\ref{dmu}) indicates that $ \mu(r) $ is always monotonically decreasing with $ r $.

\medskip

More precisely, large positive values of the chemical potential at the origin $ \nu_0 \gg 1 $ 
correspond to the degenerate 
fermions limit which is the extreme quantum case and oppositely, $ \nu_0 \ll -1 $ gives 
 the diluted limit which is the classical limit. In this classical limit the Thomas-Fermi equations
(\ref{pois})-(\ref{den}) become the equations for a self-gravitating Boltzmann gas.

We display in fig. \ref{deg} the density and velocity profiles. Namely,
we plot $ \rho(r)/\rho_0 $ and $ \sigma(r)/\sigma(0) $ as functions of 
$ r/R $ for $ \nu_0 \equiv \nu(0) = -5, \; 0, \; 5, \; 15, \;  25 $ and 
in the degenerate fermion limit $ \nu_0 \to +\infty $.
The obtained fermion profiles are always cored. 
The sizes of the cores $ r_h $ defined by eq.(\ref{onequarter}) 
are in agreement with the observations, from the compact galaxies where $ r_h \sim 35 $ pc till
the spiral and elliptical galaxies where $ r_h \sim 0.2 - 60 $ kpc. The larger and positive is 
$ \nu_0 $, the smaller is the core. The minimal core size arises in
the degenerate case  $ \nu_0 \to +\infty $ (compact dwarf galaxies).

\medskip

Eqs.(\ref{gorda})-(\ref{cero}) cover the {\bf full range} of physical galaxy situations
from the quantum degenerate fermions (dwarf galaxies) until the dilute classical limit 
(spiral and elliptic galaxies) as we discuss in subsection \ref{thofe}. 
In addition, the galaxy length scale $ l_0 $ and the galaxy mass $ M_R $ emerging from the Thomas-Fermi
approach turn to be of the same  order of magnitude of the Jeans' length and  Jeans' mass, respectively
which are here of semi-classical nature, containing $ \hbar $ and $ G $ \cite{nos}.

\subsection{Quantum degenerate limit: compact dwarf galaxies and minimal galaxy mass}

The quantum degenerate limit is very instructive and we consider it now. In this limit $ \nu_0 \gg 1 $,
the distribution function $ \Psi(E/E_0) $ becomes a step function $ \theta(-E) $ and all fermion 
states get filled till the Fermi level. Notice that the degenerate limit of the distribution function
is {\bf always} a step function $ \theta(-E) $ irrespective of the shape of the non-degenerate fermion
distribution function $ \Psi(E/E_0) $. Then, from eq.(\ref{dfI})
\be\label{Ideg}
I_n(\nu)_{deg} = \nu^{\frac{n+1}2} \; ,
\ee
The Thomas--Fermi equation (\ref{nu}) becomes 
\be
\frac{d^2 \eta}{d\xi^2} + \frac2{\xi} \; \frac{d \eta}{d\xi} = -{\eta}^{3/2} \; , \quad 
  \eta(\xi) \equiv \frac{\nu(\xi)} {\nu_0} \;  , \quad    \eta(0) = 1 \; , \quad \eta'(0) = 0  \; ,
\ee
which we solve numerically  for $ 0 < \xi < \xi_R $ with $ \xi_R =
3.6537446 $ the first zero of $ \eta(\xi) $ and the dimensionless core radius
$ \xi_h = 2.269587 $. One then finds for the mass density and total
mass in the degenerate limit 
\bea
&& \rho_{deg}(r) = \rho_0 \;\eta^{3/2}(\xi)  = \frac{2^{3/2} \; m^{5/2}}{3 \, \pi^2 \; \hbar^3} \; \mu^{3/2}(r)
\quad ,  \quad
M_{deg, \, R} = 4 \, \pi \; \rho_0 \; l_0^3 \; \int_0^{\xi_R}
    d\tau \; \tau^2 \; \eta^{3/2}(\tau) = \cr \cr \cr
&& = \left(\hbar \; \sqrt{\frac{2 \, \pi}{G}}\right)^{\! \! 3} \; 
    \frac{3 \; \sqrt{\rho_0}}{(2 \, m)^4} \; \xi_R^2 \; |\eta'(\xi_R)| 
    = 2.714 \; M_0 \, \left(\frac{{\rm keV}}{m}\right)^{\! \! 4} \,
    \sqrt{\rho_0 \, \frac{{\rm pc}^3}{M_\odot}} \; , \cr \cr
&& M_{deg, \, h} = 1.856 \; M_0 \, \left(\frac{{\rm keV}}{m}\right)^{\! \! 4} \,
    \sqrt{\rho_0 \, \frac{{\rm pc}^3}{M_\odot}} \; ,
\eea
where the numerical result  $ \xi_R^2 \; |\eta'(\xi_R)| = 2.71405512 $ and 
$ \xi_h^2 \; |\eta'(\xi_h)| = 1.855893 $ have been used and $ M_0 $ 
is given by eq.(\ref{cero}). The density in the degenerate limit vanishes where $ \mu(r) $ does.

\medskip

Finally, for the velocity dispersion we find from eqs.~(\ref{gorda}), (\ref{cero}) and (\ref{Ideg})
\be \label{sigdeg}
 \sigma_{deg}(r) = \sigma_{deg}(0) \; \eta^{1/2}(\xi) \quad , \quad 
\sigma_{deg}(0) =V_0 \, \left(\frac{{\rm keV}}{m}\right)^{\! \! \frac43}
  \left(\rho_0 \; \frac{{\rm pc}^3}{M_\odot}\right)^{\! \! \frac13} \quad  .
\ee
The phase--space density at
the origin $ Q(0) $ takes in the degenerate limit its maximal value
\be
Q_{deg}(0) = \frac{\rho_0}{\sigma^3(0)} = \frac{5\sqrt{5}}{3\pi^2} \; \frac{m^4}{\hbar^3}
= 0.3776 \; \frac{m^4}{\hbar^3} \; .
\ee
In the quantum degenerate limit $ \nu_0 \to +\infty $, the DM fermions of the halo are in the 
minimally excited state, namely the  
(semiclassical) ground state for which the three basic quantities: halo radius, galaxy mass 
$ M_h $ and velocity dispersion, take their {\it minimum} values:
\bea\label{minimo}
&& r_{h,min} = 24.51 \; \left(\frac{\rm keV}{m}\right)^{\! \! \frac43} \; 
\left(\rho_0 \; \frac{{\rm pc}^3 }{M_\odot}\right)^{\! \! -\frac16} \; {\rm pc} \quad , \quad 
\sigma_{min}(0) =  2.751 \;  \left(\frac{\rm keV}{m}\right)^{\! \! \frac43} \; 
\left(\rho_0 \; \frac{{\rm pc}^3 }{M_\odot}\right)^{\! \! \frac13} \; \frac{\rm km}{\rm s} \; ,\cr \cr
&& M_{h,min} = 1.527 \; 10^4 \; \left(\frac{\rm keV}{m}\right)^{\! \! 4} \; 
\sqrt{\rho_0 \; \frac{{\rm pc}^3}{M_\odot}} \; M_\odot \; .\label{masam}
\eea
These minimum values are similar to the estimates for degenerate fermions eqs.(\ref{semic}), as
it must be.

\medskip

The masses of compact dwarf galaxies dominated by DM must be larger than the minimum mass 
$ M_{h,min} $ eq.(\ref{masam}). The lightest known galaxy of this kind is Willman I (see Table \ref{pgal}).
Imposing $ M_{h,min} < M_{Willman ~ I} =  2.9 \; 10^4 \; M_\odot $ gives a minimal mass (a lower bound) for the WDM
particle:
\be\label{will}
m >  m_{min} = 1.91 \; {\rm keV} \; . 
\ee
It must be stressed that these minimum physical values for degenerate WDM fermions eqs.(\ref{minimo})-(\ref{will}) are
independent of the WDM particle physics model. They are 
{\bf universal} whatever is the shape of the distribution function $ f(E) $ in the non-degenerate regime.

\medskip

As can be seen from eq.(\ref{masam}) the minimal value of the WDM particle mass $ m_{min} $ 
in eq.(\ref{will}) is given by
$$
 m_{min} = 1.977 \; {\rm keV} \; \left(\frac{10^4 \; M_\odot}{M^{lightest \, galaxy}}\right)^{\! \! \frac14}
\; \left(\frac{\rho_0^{lightest \, galaxy} \; {\rm pc}^3}{M_\odot}\right)^{\! \! \frac18} \; .
$$
As a consequence, the error on $ m_{min} $ due to the observational error on the mass and central density
of the lightest  galaxy (Willman I, so far) is considerably reduced by the small exponents $ 1/4 $ and
$ 1/8 $, respectively.

\medskip

X-ray galaxy observations provide upper bounds for the particle mass for a specific WDM candidate:
the sterile neutrino. If the sterile neutrinos are described by the Dodelson-Widrow (DW) model this upper
bound turns to be \cite{casey}
\be\label{DW}
m < 2.2 \; {\rm keV} \; . 
\ee
Therefore, for WDM formed by DW sterile neutrinos their particle mass must be in the narrow interval
$ 1.91 \; {\rm keV} < m < 2.2 \; {\rm keV} $. 

\medskip

It is not easy to robustly determine the mass of Willman I and other
ultra-compact dwarfs. These objects are very dim with absolute magnitudes fainter than $ M_V
\sim - 8 $, see the discussion in refs. \cite{gil,wp,jdsmg,simon11,wolf10,datos,bwill}.
However, for such objects as Willman 1, Segue 1 and other ultra-compact dwarfs,
there exist solid structural data as the structural masses, densities and
velocity dispersions \cite{gil,wp,jdsmg,simon11,wolf10,datos,bwill}.

\medskip

More precise data for ultracompact dwarf galaxies as Willman I will make our bound eq.(\ref{will}) more precise.

Improvements on our lower bound eq.(\ref{will}) as well as on the upper bound eq.(\ref{DW}) from more precise
galaxy data will lead to more stringent constraints on sterile neutrino WDM particle models.

\subsection{Classical dilute limit: large galaxies}

In the classical dilute limit, $ \nu \ll -1 $, the FD distribution function eq.(\ref{FD}) becomes 
the Maxwell-Boltzmann distribution and we find from eqs.(\ref{nu}),
\bea\label{limcla}
&& \Psi_{cl}(y^2 -\nu) = e^{\nu -y^2}  \quad ,  \quad I_2^{cl}(\nu) = \frac34 \; \sqrt{\pi} \; e^{\nu} 
\quad ,  \quad \quad I_4^{cl}(\nu) = \frac{15}8 \; \sqrt{\pi} \; e^{\nu} 
\; ,\\ \cr
&& \frac{d^2 \nu}{d\xi^2} + \frac2{\xi} \; \frac{d \nu}{d\xi} = -
\left(\frac34 \; \sqrt{\pi}\right)^{\! \! \frac23} \; e^{\nu}
\quad , \quad   \nu(0) = \nu_0 \quad , \quad \nu'(0) = 0 \; ,\cr \cr
&& Q_{cl}(r) = \frac1{\sqrt{ 2 \; \pi^3}}  \; e^{\nu} \quad  \frac {m^4}{\hbar^3} \; .
\eea
In this limit from eq.(\ref{limcla}), the ratio $ I_4^{cl}(\nu)/I_2^{cl}(\nu) $ becomes constant and
the average velocity $ \sigma(r) $ from eqs.(\ref{gorda}) becomes uniform.
Furthermore, the numerical resolution of eq.(\ref{nu}) or eqs.(\ref{limcla})
in the  classical $ \nu_0 \ll -1 $ regime yields
\bea
&& \xi_R = 37.11 \;  e^{-\nu_0/3} \gg 1  \quad {\rm and}  \quad -\nu'(\xi_R) =  
0.05318 \; \; e^{\nu_0/3} \ll 1  \; ,\cr \cr
&& \xi_h = 3.299 \;  e^{-\nu_0/3} \gg 1  \quad {\rm and}  \quad -\nu'(\xi_h) =
0.5578  \; e^{\nu_0/3} \ll 1 \; \; . 
\eea
Therefore, we see from eqs.(\ref{gorda}), (\ref{cero} ) that the size and mass for $ \nu_0 \ll -1 $
and fixed $ \rho_0 $ grow as
\bea
&& M_R = 60.57  \; e^{-\nu_0} \; M_0 \, \left(\frac{{\rm keV}}{m}\right)^{\! \! 4} \;
    \sqrt{\rho_0 \, \frac{{\rm pc}^3}{M_\odot}} \quad , \quad
R = 694.3 \; e^{-\nu_0/3} \; \; \left(\frac{{\rm keV}}{m}\right)^{\! \! \frac43}  \; 
  \left(\rho_0 \; \frac{{\rm pc}^3}{M_\odot}\right)^{\! \! -\frac16} \; {\rm pc} \; ,\cr \cr
&& M_h = 5.0202  \; e^{-\nu_0} \; M_0 \, \left(\frac{{\rm keV}}{m}\right)^{\! \! 4} \,
    \sqrt{\rho_0 \, \frac{{\rm pc}^3}{M_\odot}} \quad , \quad
r_h = 61.71 \; e^{-\nu_0/3} \; \; \left(\frac{{\rm keV}}{m}\right)^{\! \! \frac43}  \; 
  \left(\rho_0 \; \frac{{\rm pc}^3}{M_\odot}\right)^{\! \! -\frac16} \; {\rm pc} \; . \nonumber
\eea
We see that in the classical regime, $ \nu_0 \to -\infty , \; r_h $ and $ M_h $ can be arbitrarily large. 
Galaxies of any large mass can be obtained as solutions of the Thomas-Fermi equations (\ref{nu}).

\subsection{Physical galaxy properties from compact dwarf galaxies to large galaxies}\label{thofe}

The largest value for the phase space density corresponds to the quantum degenerate fermions limit
while the smallest values appear in the classical dilute limit. 

\begin{figure}[h]
\begin{center}
\includegraphics[width=14.cm]{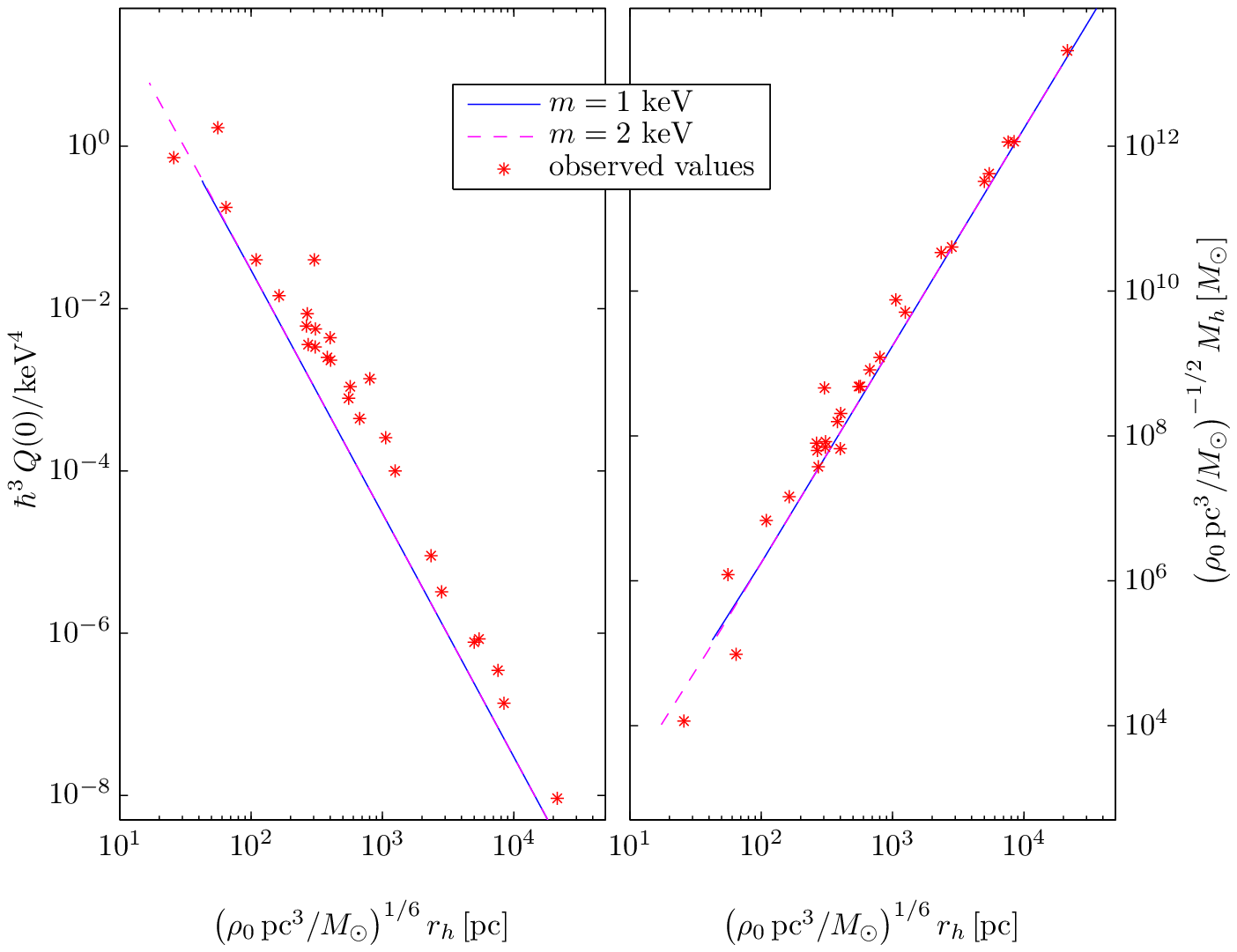}
\caption{In the left panel we display the galaxy phase-space density 
$ \hbar^3 \; Q(0)/({\rm keV})^4 =  (m/{\rm keV})^4 \;
(\sqrt{125}/3 \; \pi^2) \; [I^{\frac52}_2(\nu_0)/I_4^{\frac32}(\nu_0)] $
defined in  eq.(\ref{gorda})  obtained from the numerical resolution of the Thomas-Fermi eqs.(\ref{nu}) 
for WDM fermions of mass $ m = 1 $ and $ 2 $ keV 
versus the ordinary logarithm of the product $ \log_{10}\{r_h \; 
[{\rm pc}^3 \; \rho_0/ M_\odot]^{\frac16} \} = \log_{10}[ R_0 \; \xi_h\; ({\rm kev}/m)^\frac43] $ in parsecs. 
The red stars $ * $ are the observed values of $ \hbar^3 \; Q(0)/({\rm keV})^4 $ from Table \ref{pgal}.  
Notice that the observed values $ Q_h $ from the stars' velocity 
dispersion are in fact upper bounds for the DM $ Q_h $ and therefore the theoretical curve is slightly below them.
In the right panel we display the galaxy mass 
$ (M / M_\odot) \sqrt{M_\odot / [\rho_0 \; {\rm pc}^3]}
= 0.8230 \; 10^5 \;\left({\rm keV}/m\right)^4 \;  \xi_h^2 \;  
\left| \nu'(\xi_h) \right|/ I_2^{2/3}(\nu_0) $ obtained from the numerical resolution of the 
Thomas-Fermi eqs.(\ref{nu}) for WDM fermions of mass $ m = 1 $ and $ 2 $ keV 
versus the product $ r_h \; [{\rm pc}^3 \; \rho_0/ M_\odot]^{\frac16} = R_0 \; \xi_h \; ({\rm keV}/m)^\frac43 $ 
in parsecs.  The red stars $ * $ are the observed values of 
$ (M / M_\odot) \sqrt{M_\odot / [\rho_0 \; {\rm pc}^3]} $ 
from Table \ref{pgal}. Notice that the error bars 
of the observational data are not reported here but they are at least about $ 10-20 \%$.
Even if both masses $ m = 1 $ and 2 keV fit well the bulk of the
data, it is shown from the right panel that only $ m = 2 $ keV reproduces
the small compact dwarfs (the low mass extreme of the dashed curve). DM 
particle masses $ m $ smaller than 2 keV do not reproduce the smaller galaxy data. 
Moreover, for $ m > 2 $ keV, an overabundance of small structures appear as solution of the Thomas-Fermi 
equations which do not have observed counterpart. Therefore, $ m $ about of 2 keV is 
singled out as the most plausible value.}
\label{halo}
\end{center}
\end{figure}

\medskip

In the left panel of fig. \ref{halo} we plot from eq.(\ref{gorda}) the dimensionless quantity 
\be\label{cute1}
\frac{\hbar^3}{({\rm keV})^4} \; Q(0) = 
\frac{\sqrt{125}}{3 \; \pi^2} \; \; \frac{I^{\frac52}_2(\nu_0)}{I_4^{\frac32}(\nu_0)}
\; \left(\frac{m}{{\rm keV}}\right)^{\! \! 4} \; . 
\ee
In the right panel of fig. \ref{halo}, we plot instead the dimensionless product
\be\label{cute2}
\frac{M_h}{M_\odot} \sqrt{\frac1{\rho_0} \; \frac{M_\odot}{{\rm pc}^3}} =
 0.8230 \; 10^5 \; \frac{\xi_h^2}{I_2^{2/3}(\nu_0)} \; \; \left| \nu'(\xi_h) \right| \; \;
\left(\frac{\rm keV}{m}\right)^{\! \! 4} \; ,
\ee
where $ M_h $ is the halo mass, namely the galaxy mass inside the core radius $ r_h
$ defined by eq.(\ref{onequarter}) and we used eqs.(\ref{gorda})-(\ref{cero}).
In both cases we consider the two values $ m = 1 $ and $ 2 $ keV and we put in
the abscissa the product
\be\label{cute3}
r_h  \; \left(\frac{\rm pc^3}{M_\odot} \; \rho_0\right)^{\! \! \frac16}
= R_0 \; \xi_h  \; \left(\frac{\rm keV}{m}\right)^{\! \! \frac43}
\quad {\rm in ~ parsecs,}
\ee
where $ r_h = l_0 \; \xi_h $ is the core radius. The phase-space density $ Q(0) $ and the galaxy 
mass $ M_h $ are obtained by solving the Thomas-Fermi eqs.(\ref{nu}).
We have also superimposed the
observed values $\hbar^3\, Q_h/({\rm keV})^4 $ and $ M_h \sqrt{M_\odot / [\rho_0
  \; {\rm pc}^3]} \; \; \left(m/{\rm keV}\right)^4 $ from Table \ref{pgal}.
Notice that the observed values $ Q_h $ from the stars' velocity dispersion are
in fact upper bounds for the DM $ Q_h $. This may explain why the theoretical
Thomas-Fermi curves in the left panel of fig. \ref{halo} appear slightly below
the observational data. Notice also that the error bars of the observational
data are not reported here but they are at least about $ 10-20 \%$.

\medskip 

The phase space density decreases from its maximum value for the
compact dwarf galaxies corresponding to the limit of degenerate fermions till
its smallest value for large galaxies, spirals and ellipticals, corresponding to
the classical dilute regime. On the contrary, the halo radius $ r_h $ and the halo mass $ M_h $
 monotonically increase from the quantum to the classical regime.

Thus, the whole range of values of the chemical potential at the origin $ \nu_0
$ from the extreme quantum (degenerate) limit $ \nu_0 \gg 1 $ to the classical
(Boltzmann) dilute regime $ \nu_0 \ll -1 $ yield all masses, sizes, phase space
densities and velocities of galaxies from the ultra compact dwarfs till the
larger spirals and elliptical in agreement with the observations (see Table
\ref{pgal}).

\medskip 

From figs. \ref{halo} we can extract important information on the fermion particle WDM mass.

\medskip 

We see from the left panel fig. \ref{halo} that decreasing the DM particle mass $ m $ moves the theoretical curves
$ \hbar^3 \; Q(0) / ({\rm keV})^4 $ towards smaller $ Q(0) $ values and larger galaxy sizes, one over each other.
In the right panel of fig. \ref{halo} we see that decreasing the DM particle mass $ m $ 
displaces the theoretical curves $ (M_h / M_\odot) \sqrt{M_\odot / [\rho_0 \; {\rm pc}^3]} $ 
towards larger galaxy masses and sizes, one over each other.

The small galaxy endpoint of the curves in figs. \ref{halo}  corresponds to
the degenerate fermion limit $ \nu_0 \to +\infty $ and its value depends on the
WDM particle value $ m $. For increasing $ m $, the small galaxy endpoint moves towards
smaller sizes while for decreasing $ m $, it moves towards larger sizes.

We see from figs. \ref{halo} that decreasing the particle mass beyond a given value,
namely for particle masses $ m \lesssim 1 $ 
keV,  the theoretical curves do not reach the more compact galaxy data.
Therefore, $ m \lesssim 1 $ keV is ruled out as WDM particle mass, in agreement with
the bound eq.(\ref{will}).

For growing $ m \gtrsim $ keV the left part of the theoretical curves corresponding to the lower galaxy 
masses and sizes, will not have observed galaxy counterpart. Namely, increasing  
$ m \gg $ keV would show an overabundance of small galaxies (small scale structures) 
which do not have counterpart in the data.
This is a further indication that the WDM particle mass is approximately around 2 keV
in agreement with earlier estimations \cite{dvs,dvss}. 

\medskip

In addition, the galaxy velocity dispersions from eq.(\ref{gorda})-(\ref{cero}) turn to be fully consistent 
with the galaxy observations in Table \ref{pgal}.

\medskip

Comparing data and mass models implies the existence of a Universal DM
density distribution, may be originating from a "Universal Rotation Curve"
(URC). Spirals present universal features in their kinematics that
correlate with their global galactic properties (see \cite{pss,ps07})
which led to the discovery of the URC for spirals \cite{ps07}.
For dwarf and ellipticals galaxies, the mass modelling needs further
developpement but solid data seen to confirm the pattern shown in spirals \cite{gil,mw2},
for further discussion see Salucci in \cite{chalo14}.

\medskip

The semiclassical Thomas-Fermi approach provides much stronger results than just the quantum
bound on the phase-space density $ Q(r) $ only based on the Pauli principle,
$$
Q({\vec r}) \leq  K \; \frac{m^4}{\hbar^3} \quad , \quad K = {\cal O}(1) \; .
$$
The primordial phase-space density $ Q $ fulfils this quantum bound \cite{nos}.
Through classical time evolution, $ Q({\vec r}) $ in average  ($ \bar Q $) will always
fulfil this quantum bound since as it is known, $ \bar Q $ can only decrease with time. 
This quantum bound gives lower bounds in the galaxy core sizes in the range $ \sim 0.1 $ pc \cite{nos}
which are of the same order of magnitude as the core sizes obtained in the classical (non-quantum)
$N$-body WDM simulations \cite{mash}, which are unrealistically small cores compared with the observations.
On the contrary, the Thomas-Fermi approach includes the quantum pressure and indeed succeeds 
to provide galaxy cores with the right sizes as shown in this paper. 

\medskip

Adding baryons to CDM simulations have been often invoked to solve the
serious CDM problems at small scales. It must be noticed however that the
excess of substructures in CDM happens in DM dominated
halos where baryons are especially subdominat and hence the effects of
baryons cannot drastically modify the overabundance of substructures of the
pure CDM results.

The influence of baryon feedback into CDM cusps of density profiles
depends on the strength of the feedback. For normal values of the
feedback, baryons produce adiabatic contraction and the cusps in the density
profiles become even more cuspy.

Using the baryon feedback as a free parameter, it is possible
to exagerate the feedback such to destroy the CDM cusps
but then, the star formation ratio disagrees with the
available and precise astronomical observations.
Moreover, "semi-analytic (CDM + baryon)  models" have been introduced
which are just empirical fits and prescriptions to some galaxy observations.

In addition, there are serious evolution problems in CDM galaxies:
for instance pure-disk galaxies (bulgeless) are observed whose formation
through CDM is unexplained.

In summary, adding baryons to CDM simulations bring even more serious
discrepancies with the set of astronomical observations.

\medskip

In this paper spherical symmetry is considered for simplicity to determine 
the essential physical galaxy properties as the classical or 
quantum nature of galaxies, compact or dilute galaxies, 
the phase space density values, the
cored nature of the mass density profiles, the galaxy masses and 
sizes.  It is clear that DM halos are not perfectly 
spherical but describing them as spherically symmetric is a first
approximation to which other effects can be added.
In ref. [9] we estimated the angular momentum 
effect and this yields small corrections. The quantum or classical 
galaxy nature, the cusped or cored nature of the density profiles in the 
central halo regions can be captured in the spherically symmetric treatment.

Our spherically symmetric treatment captures the essential features
of the gravitational dynamics and agree with the observations.
Notice that we are treating the DM particles quantum mechanically through
the Thomas-Fermi approach, so that expectation values are independent
of the angles (spherical symmetry) but the particles move and fluctuate
in all directions. Namely, this is more than treating purely classical orbits
for particles in which only radial motion is present. 
The Thomas-Fermi approach can be generalized to
describe non-spherically symmetric and non-isotropic situations,
by considering  distribution functions which include other 
particle parameters like the angular momentum.

\medskip

To conclude, eqs.(\ref{gorda})-(\ref{cero}) indicate 
that the galaxy magnitudes:  halo radius, galaxy masses and velocity dispersion
obtained from the Thomas-Fermi quantum treatment for WDM fermion masses in the keV scale are
fully consistent with all the observations for all types of galaxies (see Table \ref{pgal}). 
Namely, fermionic WDM treated quantum mechanically (as it must be) is able to reproduce
the observed sizes of the DM cores of galaxies.
These results strenght the discussion in sec. \ref{qup} that compact galaxies are supported against 
gravity by the fermionic WDM quantum pressure. 

It is highly remarkably that in the context of fermionic WDM, the simple stationary
quantum description provided by the Thomas-Fermi approach is able to reproduce such broad variety of galaxies.

\medskip

In addition, WDM simulations
produce the right DM structures in agreement with observations for scales $ \gtrsim $ kpc \cite{simuw}.

\acknowledgments

We are grateful to Peter Biermann for useful discussions in many occasions.
We thank Daniel Boyanovsky and Paolo Salucci for useful remarks.

\appendix

\section{Quantum physics in simulations}

Richard P. Feynman foresaw the necessity to include quantum physics in simulations  \cite{fey}:

\medskip

{\it ``I'm not happy with all the analyses that go with just the classical theory, because nature isn't classical, dammit, and if you want to make a simulation of nature, you'd better make it quantum mechanical, and by golly it's a wonderful problem, because it doesn't look so easy.'' }

\end{document}